\documentclass[apjl,a4paper,12pt,useAMS]{emulateapj}
\usepackage{amssymb}
\usepackage{graphicx}

\begin{document}
\title{Bright ``merger-nova" from the remnant of a neutron star binary merger: A signature of a newly born, massive, millisecond magnetar }
\author{Yun-Wei Yu\altaffilmark{1},
Bing Zhang\altaffilmark{2,3},
He Gao\altaffilmark{2}}

\altaffiltext{1}{Institute of Astrophysics, Central China Normal
University, Wuhan 430079, China, {yuyw@mail.ccnu.edu.cn}}
\altaffiltext{2}{Department of Physics and Astronomy, University of
Nevada, Las Vegas, NV 89154, USA, {zhang@physics.unlv.edu}}
\altaffiltext{3}{Kavli Institute for Astronomy and Astrophysics and
Department of Astronomy, Peking University, Beijing 100871, China}

\begin{abstract}
A massive millisecond magnetar may survive the merger of a neutron
star (NS) binary, which would continuously power the merger ejecta.
We develop a generic dynamic model for the merger ejecta with energy
injection from the central magnetar. The ejecta emission (the
``merger-nova") powered by the magnetar peaks in the UV band and the
peak of the light curve progressively shifts to an earlier epoch
with increasing frequency. A magnetar-powered mergernova could have
an optical peak brightness comparable to a supernova, which is a few
tens or hundreds times brighter than the radioactive-powered
merger-novae (the so-called macro-nova or kilo-nova). On the other
hand, such a merger-nova would peak earlier and have a significantly
shorter duration than that of a supernova. An early collapse of the
magnetar could suppress the brightness of the optical emission and
shorten its duration. Such millisecond-magnetar-powered merger-novae
may be detected from NS-NS merger events without an observed short
gamma-ray burst, and could be a bright electromagnetic counterpart
for gravitational wave bursts due to NS-NS mergers. If detected, it
suggests that the merger leaves behind a massive NS, which has
important implications for the equation-of-state of nuclear matter.
\end{abstract}

\section{Introduction}
Mergers of double neutron stars (NSs) or a NS with a stellar-mass
black hole are the primary targets of direct detections of
gravitational waves (GWs). It is expected that, by the end of this
decade, the second generation of ground-based GW detectors would
extend the detection horizons of the mergers to a few hundred Mpcs
or even 1 Gpc (Abadie et al. 2010; Nissanke et al. 2013).
Electromagnetic (EM) transients that are spatially and temporally
coincident with the GW bursts due to the mergers could play a
crucial role in the discovery and identification of the GW signals
by providing position, time, redshift, and astrophysical properties
of the sources.

The brightest EM emission during the compact binary mergers is
probably short-duration gamma-ray bursts (SGRBs; e.g. Gehrels et al.
2005; Fox et al. 2005; Fong et al. 2013; cf. Virgili et al. 2011).
However, since they are usually beamed into a small opening angle
(e.g. Burrows et al. 2006; De Pasquale et al. 2010), most GW bursts
would not be detected together with SGRBs (e.g. Metzger \& Berger
2012). Numerical simulations show that a more isotropic,
sub-relativistic ($v_{\rm ej}\sim 0.15-0.25c$) outflow could be
ejected during a merger, which could include the tidal tail matter
during the merger and the matter from the accretion disk (e.g.
Rezzolla et al. 2011; Bauswein et al. 2013; Rosswog et al. 2013).
The typical mass of the ejecta is in the range $M_{\rm ej}\sim
10^{-4}-10^{-2}M_{\odot}$ (Hotokezaka et al. 2013). Somewhat higher
mass could also exist (Fan et al. 2013). The ejecta is expected to
be neutron-rich and thus heavier radioactive elements could be
synthesized via r-process. Li \& Paczynski (1998) suggested that the
ejecta could produce a thermal UV-optical transient powered by
radioactive decay, which is more isotropic than SGRBs. In the past
few years, much effort has been invested in determining the details
of merger dynamics, nuclear synthesis, radiative transfer, etc
(e.g., Kulkarni 2005; Rosswog 2005; Metzger et al. 2010; Goriely et
al. 2011; Roberts et al. 2011; Barnes \& Kasen 2013; Bauswein et al.
2013; Grossman et al. 2013; Piran et al. 2013; Rosswog et al. 2013;
Takami et al. 2013; Tanaka \& Hotokezaka 2013). The interaction
between the merger ejecta and the ambient medium is also expected to
produce a long-lasting afterglow emission (Nakar \& Piran 2011;
Metzger \& Berger 2012; Piran et al. 2013). Nevertheless, the
brightness of the afterglow emission is typically low, and strongly
depends on the ambient density.

The merger products are usually considered to be a black hole.
Alternatively, given the uncertainties of nuclear matter
equation-of-state and NS mass distributions of the merger systems,
it is possible that at least some NS-NS mergers would leave behind a
stable (for hours to days), rapidly rotating NS (e.g. Dai et al.
2006; Fan \& Xu 2006; Zhang 2013; Giacomazzo \& Perna 2013). This is
indirectly supported by the fact that some SGRBs are followed by an
X-ray plateau with an abrupt ending, which is best interpreted as
emission from a spinning down magnetar (e.g. Rowlinson et al. 2010,
2013). More directly, the present lower limit of the maximum mass of
Galactic NSs is precisely set by PSR J0348+0432 to
$2.01\pm0.04M_{\odot}$ (Antoniadis et al. 2013). The permitted
equations of state usually lead to a maximum mass close to or higher
than $2.5M_{\odot}$ for a non-rotating NS, which is comparable to
the sum of the masses of some Galactic NS-NS binaries. Zhang (2013)
proposed that if a NS-NS merger leaves behind a millisecond
magnetar, a GW burst would be associated with a bright X-ray early
afterglow due to magnetar wind dissipation, regardless of whether
there is an associated SGRB. Gao et al. (2013) studied the
multi-wavelength afterglows of such a magnetar-powered merger
ejecta.

Following the above consideration, we suggest in this Letter that
the magnetar wind would first heat up the neutron-rich merger ejecta
before powering its afterglow and consequently produce a bright
``merger-nova"\footnote{The thermal emission of the merger ejecta
(Li \& Paczynski 1998) was named as ``macro-nova" by Kulkarni (2005)
due to its sub-supernova luminosity, or as ``kilo-nova" by Metzger
et al. (2010) due to its luminosity of $\sim10^3$ times than the
Eddington luminosity. In this letter, we use a more general word
``merger-nova" to reflect a wider range of predicted luminosities.}.
A similar process could have been observed in the so-called
superluminous supernovae (Kasen \& Bildsten 2010), the light curves
of which can be explained by having a millisecond magnetar wind
heating the ejecta more than radioactive processes (Inserra et al.
2013). In this letter, we develop a dynamic model for the evolution
of the merger ejecta including acceleration, coasting, and
deceleration. We then study the emission properties of the
millisecond-magnetar-powered merger-nova and the follow-up
broad-band afterglows (see also Gao et al. 2013). These predicted EM
signals can serve as interesting targets for in search for EM
counterparts of GW burst triggers in the upcoming Advanced
LIGO/Virgo era.
%%%%%%%%%%%%%%%%%%%%%%%%%%%%%%%%%%%%%%%%%%%%%%%%%%%%%%%%%%%%%%%%%%%%

\section{Merger-nova emission}
For a millisecond magnetar of an initial spin period $P_i$, its
total rotational energy reads $E_{\rm
rot}=2\times10^{52}P_{i,-3}^{-2}~ \rm erg$. Hereafter the convention
$Q_x=Q/10^x$ is adopted in cgs units. With a dipolar magnetic field
of strength $B$, the spin-down luminosity of the millisecond
magnetar as a function of time may be expressed by the magnetic
dipole radiation formula
\begin{eqnarray}
L_{\rm sd}=L_{\rm sd,i}\left(1+{t\over t_{\rm
md}}\right)^{-2}\label{Lp}
\end{eqnarray}
with $L_{\rm sd,i}=10^{47}~R_{s,6}^6B_{14}^{2}P_{i,-3}^{-4}\rm
~erg~s^{-1}$ and $t_{\rm
md}=2\times10^{5}~R_{s,6}^{-6}B_{14}^{-2}P_{i,-3}^{2}\rm s$, where
$t$ is time in the observer's frame. Based on the derived magnetar
parameters by fitting the SGRB X-ray plateau feature (Rowlinson et
al. 2013) and fitting the superluminous supernovae (Inserra et al.
2013), we adopt the following parameters as reference values:
$B_{14}=5$ and $P_{i,-3}=5$, which give $L_{\rm
sd,i}=1.2\times10^{46}\rm ~erg~s^{-1}$ and $t_{\rm
md}=6.7\times10^{4}$ s, where a relatively large stellar radius
$R_{s,6}=1.2$ is adopted by considering a rapidly rotating
supra-massive NS.

The magnetar wind runs into the essentially isotropic ejecta with an
ultra-relativistic speed and is quickly decelerated by the ejecta.
At the same time, the injected wind continuously pushes from behind
and accelerate the ejecta. A forward shock crosses the ejecta within
seconds (Gao et al. 2013), after which it propagates into the
interstellar medium. Since the ejecta is ``sandwiched'' between the
forward shock and the magnetar wind, its internal energy steadily
increases (heats up) as the forward shock speed increases. The wind
energy could be deposited into the ejecta either via direct energy
injection by a Poynting flux (Bucciantini et al. 2012), or due to
heating from the bottom by the photons generated in a dissipating
magnetar wind via forced reconnection, as the wind is decelerated by
the ejecta (e.g. Zhang 2013). The latter process can be efficient,
which would give rise to a relatively large efficiency $\xi$ (for
which we take a nominal value 0.3, Zhang \& Yan 2011) of injecting
spin-down luminosity to the ejecta. The heating process continues
until the supra-massive magnetar collapses to a black hole at
$t_{\rm col}$, which could not be much longer than $t_{\rm md}$
after which the star remarkably is spun down and loses a significant
centrifugal support. In principle, a smaller collapsing time could
be expected if some other spin-down mechanisms (e.g., gravitational
radiation) act earlier.

An obvious difference between a merger-nova and a supernova would be
their distinct ejecta masses. We use $E_{\rm rot}/M_{\rm ej}c^2=1$
to define a critical ejecta mass as $M_{\rm
ej,cr}=0.01P_{i,-3}^{-2}M_{\odot}$, below which the ejecta can be
accelerated to a relativistic speed\footnote{The impact of a
magnetar wind on a merger-nova was considered by Kulkarni (2005),
who adopted a relatively long spin period of a few hundred
milliseconds, so that the dynamics is still in the non-relativistic
regime.} by the magnetar wind (Gao et al. 2013). The predicted range
of the ejecta masses, $M_{\rm ej} \sim 10^{-4}-10^{-2} M_\odot$,
indicates that a complete description of the dynamical evolution of
a merger-nova, which covers both the non-relativistic and
high-relativistic phases, is desirable. The total energy of the
ejecta excluding the rest energy can be expressed by $E_{\rm
ej}=(\Gamma-1)M_{\rm ej}c^2+\Gamma E'_{\rm int}$, where $\Gamma$ is
the Lorentz factor and $E'_{\rm int}$ is the internal energy
measured in the comoving rest frame. Energy conservation gives
$dE_{\rm ej}=\left(\xi L_{\rm sd}+L_{\rm ra}-L_{\rm e}\right)dt$,
where a fraction $\xi$ of the spin-down luminosity is assumed to be
injected into the ejecta, $L_{\rm ra}$ is the radioactive power, and
$L_{\rm e}$ is the radiated bolometric luminosity. The dynamic
evolution of the ejecta can be determined by
\begin{eqnarray}
{d\Gamma\over dt}={\xi L_{\rm sd}+L_{\rm ra}-L_e-\Gamma {\cal
D}({dE'_{\rm int}/ dt')}\over M_{\rm ej}c^2+E'_{\rm
int}},\label{Gamma}
\end{eqnarray}
where ${\cal D}=1/[\Gamma(1-\beta)]$ is the Doppler factor with
$\beta=\sqrt{1-\Gamma^{-2}}$. The comoving time $dt'$ can be
connected with the observer's time by $dt'={\cal D}dt$. The
variation of the internal energy in the comoving frame can be
expressed by (e.g. Kasen \& Bildsten 2010)
\begin{eqnarray}
{dE'_{\rm int}\over dt'}=\xi L'_{\rm sd}+ L'_{\rm ra} -L'_{\rm e}
-\mathcal P'{dV'\over dt'},\label{Eint}
\end{eqnarray}
where the comoving luminosities read $L'_{\rm sd}=L_{\rm sd}/{\cal
D}^2$, $L'_{\rm e}=L_{\rm e}/{\cal D}^2$, and
\begin{eqnarray}
L'_{\rm ra}&=& {L_{\rm ra}\over {\cal
D}^2}\nonumber\\
&=&4\times10^{49}M_{\rm
ej,-2}\left[{1\over2}-{1\over\pi}\arctan \left({t'-t'_0\over
t'_\sigma}\right)\right]^{1.3}~\rm erg~s^{-1}
\end{eqnarray}
with $t'_0 \sim 1.3$ s and $t'_\sigma \sim 0.11$ s (Korobkin et al.
2012). For typical parameters $L_{\rm ra}\ll L_{\rm sd}$ unless the
magnetar only lives for a few hundred seconds. $\mathcal P'dV'$
represents the work due to free expansion of the ejecta which
converts internal energy into bulk kinetic energy. The pressure
$\mathcal P'=E'_{\rm int}/3V'$ is dominated by radiation, and the
evolution of the comoving volume can be determined by
\begin{eqnarray}
{dV'\over dt'}=4\pi R^2\beta c,
\label{Vs}
\end{eqnarray}
together with
\begin{eqnarray}
{dR\over dt}={\beta c\over (1-\beta)},\label{R}
\end{eqnarray}
where $R$ is the radius of the ejecta. The radiated bolometric
luminosity can be derived approximately from the diffusion equation
in the comoving frame (Kasen \& Bildsten 2010; Kotera et al. 2013)
\begin{eqnarray}
L'_e&=&{E'_{\rm int}c\over \tau R/\Gamma}={E'_{\rm int}t'\over {t'_{\rm d}}^2},{\rm ~for~} t\leq t_\tau,\nonumber\\
&=&{E'_{\rm int}c\over R/\Gamma},{\rm ~~~~~~~~~~~~for~}
t>t_\tau,\label{Le}
\end{eqnarray}
where $\tau=\kappa (M_{\rm ej}/V')(R/\Gamma)$ is the optical depth
of the ejecta with $\kappa$ being the opacity\footnote{In our
calculations, a constant opacity $\kappa=0.2~\rm cm^2g^{-1}$ is
adopted for simplicity, which is appropriate for electron scattering
in a plasma with an ionization degree of 0.5. However, for r-process
elements, Kasen et al. (2013) found that the bound-bound,
bound-free, and free-free transitions could provide more important
contributions to the opacity, which makes the opacity higher and
strongly energy-dependent. As a result, the merger-nova emission
could be extended, weakened, and shifted towards softer bands
(Barnes \& Kasen 2013). Additionally, the ionization of the ejecta
by the wind X-ray emission (Zhang 2013) could also affect the
opacity.}, $t'_{\rm d}\equiv\left(\tau R t'/\Gamma c\right)^{1/2}$
is the effective diffusion time, and $t_\tau$ is the time at which
$\tau=1$. We note that the optical depth reads $\tau=\Gamma
ct'/R\approx \beta^{-1}>1$ when $t=t_{\rm d}$, which suggests that
$t_{\tau}>t_{\rm d}$ all the time, similar to the non-relativistic
case (Kasen \& Bildsten 2010).

\begin{figure}
\centering\resizebox{0.8\hsize}{!}{\includegraphics{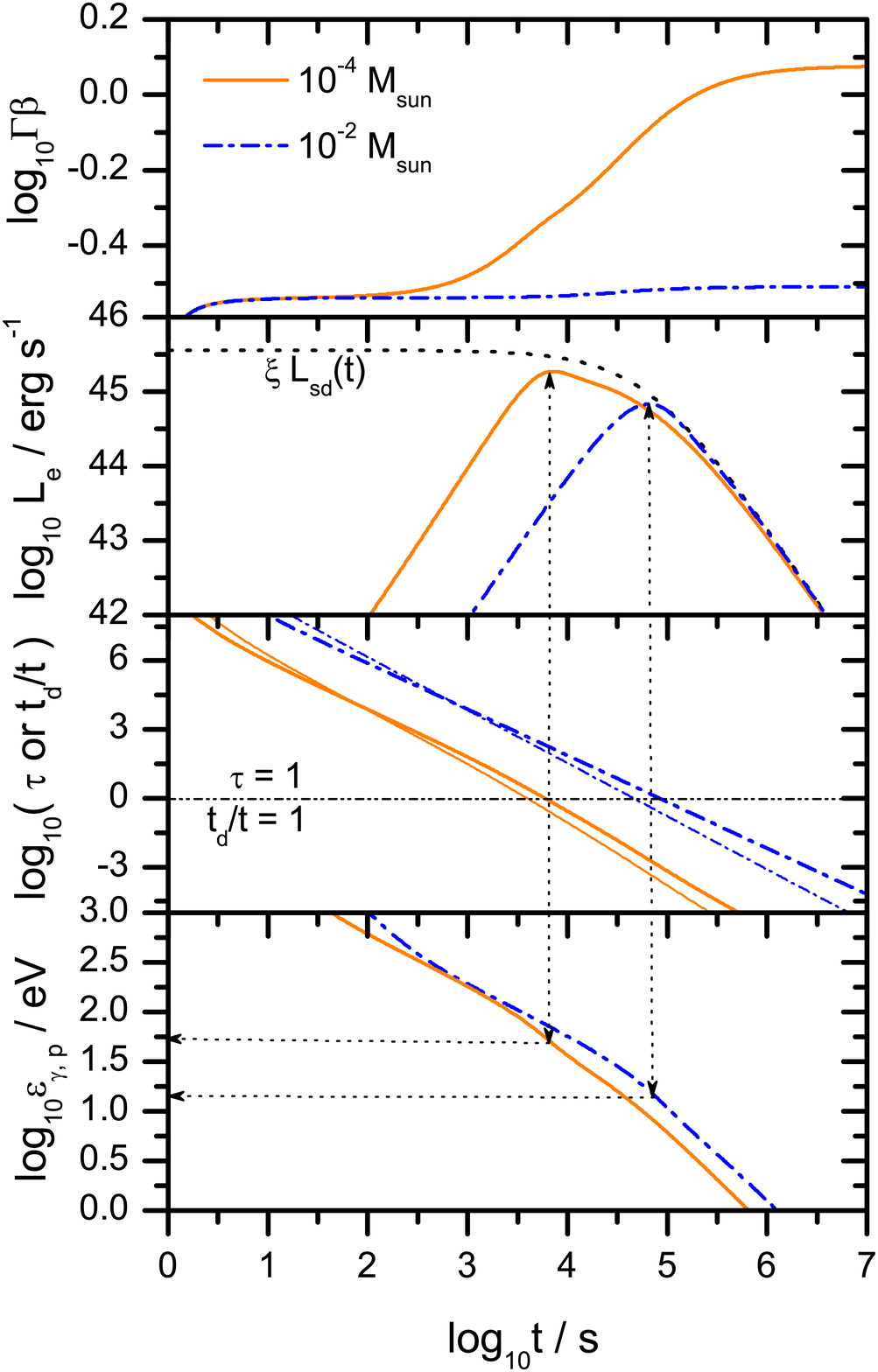}}
\caption{From top to bottom, evolutions of dynamics, bolometric
luminosity, optical depth (thick) and ratio $t'_{\rm d}/t'$ (thin),
as well as black body peak energy of merger-novae with two typical
ejecta masses: $M_{\rm ej}= 10^{-4}M_{\odot}$ (solid) and
$10^{-2}M_{\odot}$ (dash-dotted). The dotted arrows indicate that
(i) the peak time of the merger-nova emission locates at $\tau=1$
(for $10^{-4}M_{\odot}$) or $t_{\rm md}$ (for $10^{-2}M_{\odot}$)
and (ii) the peak energy corresponding to the peak luminosity is in
the range of $15-50$ eV. The initial velocity of the ejecta is taken
$\beta_i=0.2$ and the initial ejecta energy $E'_{\rm
int,i}=E_{k,i}={(1/2)M_{\rm ej}\beta_i^2c^2}$. The magnetar
parameters are: $B_{14}=5$, $P_{i,-3}=5$, and $\xi=0.3$. }
\end{figure}

Numerical solutions to the above equations are presented in Fig. 1
for two ejecta masses, $10^{-2} M_\odot$ and $10^{-4} M_\odot$,
where the magnetar collapse effect is not included. For the low mass
case, the dynamical transition from the non-relativistic regime to
the mildly-relativistic regime is clearly shown in the top panel.
The acceleration time of the ejecta is determined by the spin-down
timescale $t_{\rm md}$ in all the situations. The optical depth and
the diffusion time play a crucial role in determining the temporal
behavior of the merger-nova emission. As $\tau$ and the ratio
$t_{\rm d}/t$ gradually drop towards unity, the bolometric light
curve rises and finally reaches a peak at $t_{\rm d}$, $t_{\tau}$,
or $t_{\rm md}$ (i.e. the ejecta photosphere). To be specific, for a
high-mass ejecta, the peak time
%\footnote{\textbf{Strictly, the peak
%could occur a factor of a few in time before $t_{\rm d}$ (Dexter \&
%Kasen 2013).}}
$t_{\rm peak}\sim t_{\rm d}$ for $t_{\rm md}<t_{\rm d}<t_{\tau}$
(e.g., the case of superluminous supernova) and $t_{\rm peak}\sim
t_{\rm md}$ for $t_{\rm d}<t_{\rm md}<t_{\tau}$ (e.g., the case of
$M_{\rm ej}=0.01M_{\odot}$ in Fig. 1), which were analytically
proved by Kasen \& Bildsten (2010) and Dexter \& Kasen (2013). In
contrast, since a low-mass ejecta could become optically thin before
$t_{\rm md}$, the emission would monotonously decrease after
$t_{\tau}$ even though there is further energy injection. So for
$t_{\rm d}<t_{\tau}<t_{\rm md}$ (e.g., the case of $M_{\rm
ej}=10^{-4}M_{\odot}$ in Fig. 1), the luminosity peak could appear
at $t_{\tau}$. Fig. 1 shows that the peak time of merger-novae could
range from hours (low $M_{\rm ej}$ case) to $\sim$ a day (high
$M_{\rm ej}$ case). After $t_{\rm peak}$, the bolometric luminosity
starts to decrease and the decrease rate approximately track the
spin-down luminosity at $t>{\rm max} (t_\tau, t_{\rm d}, t_{\rm
md})$.

\begin{figure}
\resizebox{\hsize}{!}{\includegraphics{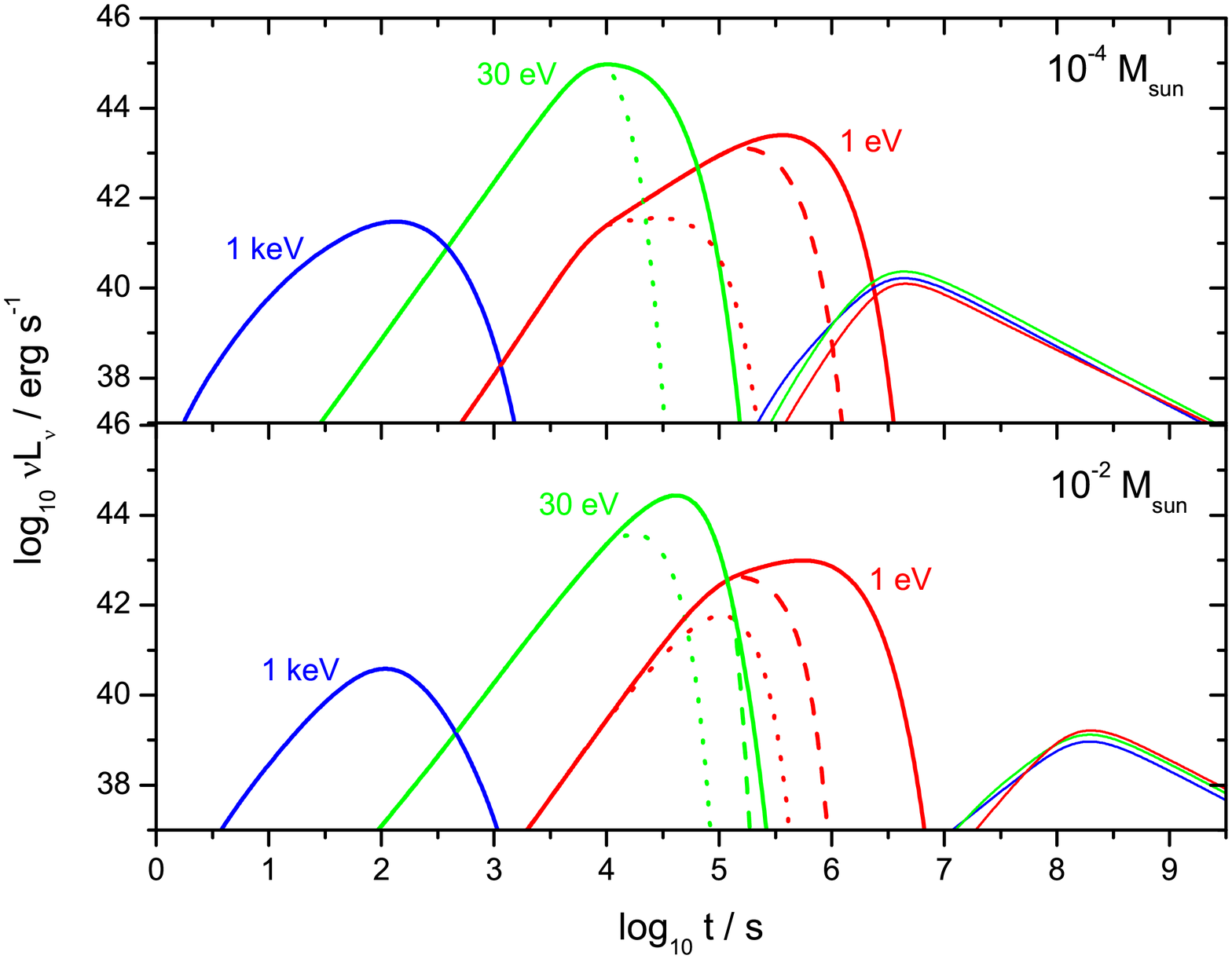}}
\centering\caption{Light curves of the merger-nova (thick) and
afterglow (thin) emissions at different observational frequencies as
labeled. The dashed and dotted lines are obtained for an optionally
taken magnetar collapsing time as $t_{\rm col}=2t_{\rm md}$ and
$t_{\rm col}=10^4$ s, respectively. The ambient density is taken as
$0.1~\rm cm^{-3}$, and other model parameters are the same as Figure
1.}
\end{figure}

The peak energy of the emission spectrum $\nu L_{\nu}$ can be
characterized by the blackbody temperature, specifically
\begin{eqnarray}
\varepsilon_{\gamma,p}\approx 4{\cal D}kT'=4{\cal D}k\left({E'_{\rm int}\over
aV'}\right)^{1/4},
\end{eqnarray}
where $k$ is the Boltzmann constant and $a$ the radiation constant.
As shown in the bottom panel of Fig. 1, the blackbody peak of the
photosphere emission mostly falls into the $\sim 15-50$ eV energy
band for the adopted parameters. For an observational frequency
$\nu$, the luminosity light curve can be calculated as
\begin{eqnarray}
\nu L_{\nu}={1\over\max(\tau,1)}{8\pi^2  {\cal D}^2R^2\over
h^3c^2}{(h\nu/{\cal D})^4\over \exp(h\nu/{\cal D}kT')-1},
\end{eqnarray}
where $h$ is the Planck constant. The light curves at different
frequencies (1 eV, 30 eV, 1 keV) of the millisecond-magnetar-powered
merger-novae are presented in Fig. 2. For nominal parameters, the
emission mainly occurs in the UV band with a peak luminosity around
$10^{45}\rm erg~s^{-1}$. Higher-energy (e.g. X-ray) emission peaks
earlier and the corresponding luminosity decreases significantly
(due to the exponential tail of thermal emission) with increasing
photon energy. In the optical band ($\sim1$ eV), a luminous flash
with a peak luminosity of $\sim 10^{43}\rm erg~s^{-1}$ appears in
the day to week time scale. This was the reason why we did not adopt
the word ``macro-nova" or ``kilo-nova". Nevertheless, such a bright
optical emission could be significantly suppressed by an early
collapse of the magnetar ($t_{\rm col}\ll t_{\rm md}$) due to an
extra angular momentum loss (e.g. via strong gravitational
radiation), as shown by the dotted lines in Fig. 2 for an optionally
taken $t_{\rm col}=10^4$ s. Of course, in a more detailed
calculation, the influence on the spin down behavior of the extra
angular momentum loss before this collapsing time should also be
taken into account (Fan et al. 2013). For a direct impression of the
merger-nova optical emission, in Figure 3 we present the optical
light curve of the magnetar-powered merger-nova in linear time
scale, in comparison with the bolometric light curves of two
supernovae (SN 1998bw and SN 2006gy) and a light curve of
radioactive merger-nova (Eqs. \ref{Gamma} and \ref{Eint} without the
magnetar term). As shown, the lifetime of the magnetar plays a
crucial role to determine the brightness and duration of the
merger-nova optical emission.

\begin{figure}
\resizebox{\hsize}{!}{\includegraphics{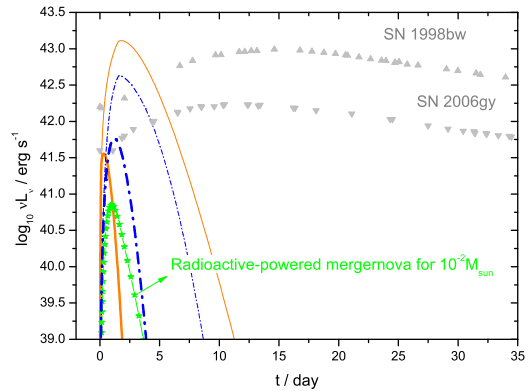}} \caption{Optical
($\sim1$ eV) light curves of the millisecond-magnetar-powered
merger-nova, in comparison with the light curves of two supernovae
(bolometric) and one radioactive-powered merger-nova (as labeled).
The dash-dotted (blue) and solid (orange) lines represent $M_{\rm
ej}=10^{-2}M_{\odot}$ and $10^{-4}M_{\odot}$, respectively. The
thick and thin lines correspond to a magnetar collapsing time as
$t_{\rm col}=10^4s\ll t_{\rm md}$ and $t_{\rm col}=2t_{\rm md}$,
respectively. The zero-times of the supernovae are set at the first
available data.}
\end{figure}

\begin{figure}
\centering\resizebox{\hsize}{!}{\includegraphics{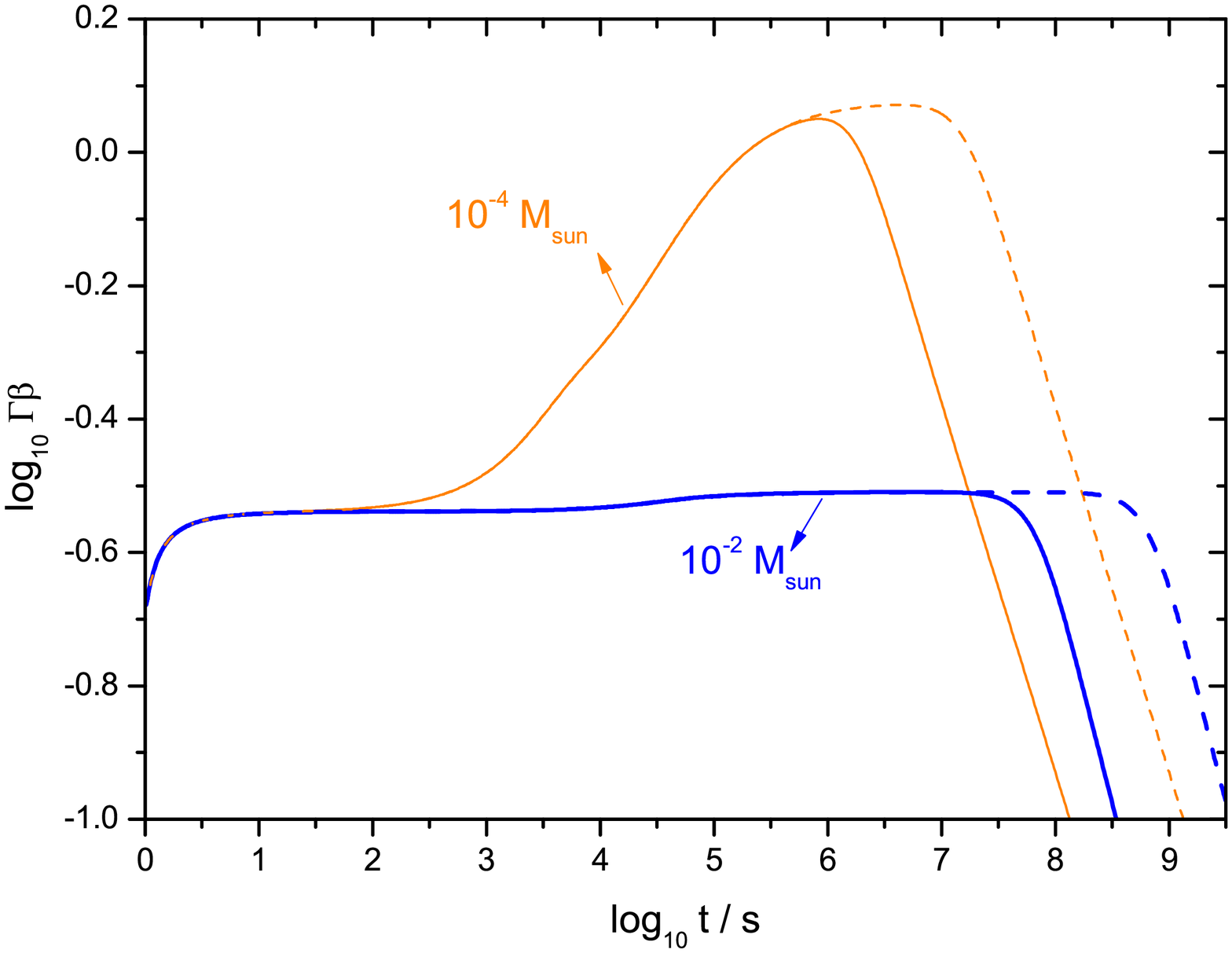}}
\caption{Dynamic evolutions of the millisecond-magnetar-powered
merger-novae and their afterglows for the ambient density
$n=1\rm~cm^{-3}$ (solid) and $10^{-3}\rm cm^{-3}$ (dashed),
respectively.}
\end{figure}
\section{Afterglow from external shock}
For a full dynamical description of the system, here we consider the
deceleration of the merger ejecta by sweeping up the ambient medium
(see also Gao et al. 2013). The treatment is similar to the generic
dynamic model for GRB afterglow (Huang et al. 1999), but with
continuous energy injection from the magnetar (Dai \& Lu 1998a,b;
Zhang \& M\'esz\'aros 2001). The total energy of the ejecta and
shocked medium can be expressed as $E=(\Gamma-1)M_{\rm ej}c^2+\Gamma
E'_{\rm int}+(\Gamma^2-1)M_{\rm sw}c^2$, where $M_{\rm sw}$ is the
mass of the swept up medium, and the comoving internal energy of the
shocked medium is $(\Gamma-1) M_{\rm sw}c^2$ according to the shock
jump condition. The energy conservation law gives
\begin{eqnarray}
{d\Gamma\over dt}={\xi L_{\rm sd}+L_{\rm ra}-L_{\rm e}-\Gamma {\cal
D}\left({dE'_{\rm int}\over
dt'}\right)-(\Gamma^2-1)c^2\left({dM_{\rm sw}\over dt}\right)\over
M_{\rm ej}c^2+E'_{\rm int}+2\Gamma M_{\rm sw}c^2}\label{Gamma2}
\end{eqnarray}
where the energy loss due to shock emission is ignored, an
approximation usually adopted in GRB afterglow modeling. As shown in
Figure 4, for a reasonable range of the ambient density,
deceleration could not start before acceleration is completed.
Therefore, the acceleration and deceleration processes can be in
principle investigated independently, as treated in Section 2. The
light curves of the afterglow synchrotron emission for a typical
ambient density $n=0.1\rm cm^{-3}$ are presented in Figure 2 along
with the merger-nova light curves. As shown, the afterglow emission
could be much weaker than that of the merger-nova in a wide
frequency range, although a noteworthy fraction of the injected
energy is also transferred to the shock.

\section{Conclusion and discussion}

By describing the dynamic evolution of a merger ejecta powered by a
millisecond magnetar, we calculate the thermal emission of the
merger-nova and the non-thermal emission of the external shock. The
optical brightness of the millisecond-magnetar-powered merger-nova
is found to be comparable to or even higher than that of supernovae,
which is a few tens or hundreds times brighter than the
radioactive-powered kilo-novae, if the magnetar remains stable
before $t_{\rm md}$. Nevertheless, early GW loss and an earlier
collapsing time could suppress the optical emission significantly.
The magnetar collapse due to losing most centrifugal support could
also restrict the duration of the mergernova within the order of (at
most) a few days, which is considerably shorter than the supernovae
duration lasting months and years. Detecting such a unique EM
transient associated with a GW burst would unambiguously confirm the
astrophysical origin of the GW burst and robustly suggest a massive
millisecond magnetar formed during the merger.

So far, no bright optical merger-nova was detected in association
with SGRBs. This may be understood as follows. Along the spin axis,
a strong magnetar jet could breaks out by propelling ejecta sideways
(Bucciantini et al. 2012; Quataert \& Kasen 2012), so that there
could be no merger-nova emission toward the observer in the SGRB
direction. A bright merger-nova may still be observable in the
equatorial direction, but it is relativistically Doppler de-boosted
in the direction of the SGRB. We expect that bright merger-nova tend
to be discovered in NS-NS mergers without a SGRB association.

%A detailed Monte Carlo simulation on the NS-NS merger event rate,
%redshift distribution, the fraction that harbors a stable magnetar,
%quantities of the post-merger system, as well as the telescope's
%survey modes and sensitivities, is needed to give a reliable
%estimate about the detectability of these events.

For the detectability of the millisecond-magnetar-powered
mergernovae, a detailed Monte Carlo simulation could be desirable.
Here we give a rough estimate. In the survey mode, the detection
efficiency of merger-novae by an optical telescope may be estimated
by
\begin{eqnarray}
\eta\sim{(1+z)T_{\rm mn}\over T_{\rm exp}}{\rm FOV\over4\pi}\sim
2\times10^{-4}(1+z),
\end{eqnarray}
where $z$ is redshift, $T_{\rm mn}\sim \rm days$ is the merger-nova
duration above the detector sensitivity limit, $T_{\rm exp}\sim\rm
hours$ is the exposure time, $\rm FOV\sim10^{-4}$ is the field of
view of the telescope. The detection rate of
millisecond-magnetar-powered bright optical merger-novae without a
SGRB association may be estimated as
\begin{eqnarray}
{\cal R}_{\rm mn} &=& \dot{\rho}_{_{\rm NS-NS}}{4\pi\over
3}\left[{L_{\rm mn}\over 4\pi S(1+z)^2}\right]^{3/2}\eta f\nonumber\\
&=&0.06 (1+z)^{-2}L_{\rm mn,
42.7}^{3/2}S_{-13.7}^{-3/2}\eta_{-4}f_{-1}\nonumber\\
&&\times\left({\dot{\rho}_{_{\rm NS-NS}}\over 500\rm
Gpc^{-3}yr^{-1}}\right)\rm yr^{-1},
\end{eqnarray}
where $\dot \rho_{_{\rm NS-NS}}$ is the NS-NS merger rate density
normalized to beaming-corrected SGRB rate density, $L_{\rm mn}$ is
the merger-nova luminosity, $S$ is the telescope sensitivity which
is normalized to a V-band magnitude $\sim22.5$ and should strongly
depend on the exposure time, and $f$ is the fraction of NS-NS
mergers that give rise to a millisecond magnetar. One can see that
the short duration of the merger-novae (small $\eta$) could make
them easily to evade from the current supernova surveys, even though
they are very luminous. A shorter lifetime of the magnetar (e.g. due
to strong gravitational radiation) would reduce $L_{\rm mn}$ and
$T_{\rm mn}$, which lead to a lower observed event rate of the
optical merger-novae. Future wide-field optical telescope surveys
(e.g. the Ground-based Wide-Angle Camera array, GWAC) would detect
these events or pose important constraints on the unknown parameters
such as $\dot \rho_{_{\rm NS-NS}}$, $\eta$, and $f$.

Finally, while this Letter only focuses on the effect of
energy injection into the merger ejecta, a large
fraction of the spin-down luminosity carried in the
magnetar wind could be dissipated directly. The internal dissipation may
arise from turbulent magnetic reconnection due to internal
collisions of the magnetar wind (Zhang \& Yan 2011) or from the
termination shock of the wind (Dai 2004). These internal
dissipations could produce an emission typically in a higher energy
band, e.g. in X-rays (Yu et al. 2010; Zhang 2013). This X-ray
transient is also expected to ionize the entire ejecta, similar to
the case of superluminous supernovae (Metzger et al. 2013).

\acknowledgements The authors acknowledge the anonymous referee for
helpful suggestions, N. Bucciantini, Z.-G. Dai, Y.-Z. Fan, L.-X. Li
and R.-X. Xu for useful comments and discussions, and C. Inserra for
providing the supernova data. This work is supported by the 973
program (Grant No. 2014CB845800), the National Natural Science
Foundation of China (Grant No. 11103004), and the Funding for the
Authors of National Excellent Doctoral Dissertations of China (Grant
No. 201225).
%BZ is supported by NSF AST-0908362.

\end{document}